\documentclass[pra,twocolumn,showpacs,preprintnumbers,amsmath,amssymb,article,superscriptaddress]{revtex4-1}
\pdfoutput=1
\usepackage{graphicx}
\usepackage{float}
\usepackage{verbatim}
\usepackage{dcolumn}
\usepackage{bm}
\usepackage{dsfont}
\usepackage{color}

\begin{document}


\newcommand{\beq}{\begin{equation}}
\newcommand{\eeq}{\end{equation}}
\newcommand{\beqa}{\begin{eqnarray}}
\newcommand{\eeqa}{\end{eqnarray}}
\newcommand{\lf}{\hfil \break \break}
\newcommand{\ahat}{\hat{a}}
\newcommand{\adag}{\hat{a}^{\dagger}}
\newcommand{\adagg}{\hat{a}_g^{\dagger}}
\newcommand{\bhat}{\hat{b}}
\newcommand{\bdag}{\hat{b}^{\dagger}}
\newcommand{\bdagg}{\hat{b}_g^{\dagger}}
\newcommand{\chat}{\hat{c}}
\newcommand{\cdag}{\hat{c}^{\dagger}}
\newcommand{\dhat}{\hat{d}}
\newcommand{\nhat}{\hat{n}}
\newcommand{\ndag}{\hat{n}^{\dagger}}
\newcommand{\den}{\hat{\rho}}
\newcommand{\phihat}{\hat{\phi}}
\newcommand{\Ahat}{\hat{A}}
\newcommand{\Adag}{\hat{A}^{\dagger}}
\newcommand{\Bhat}{\hat{B}}
\newcommand{\Bdag}{\hat{B}^{\dagger}}
\newcommand{\Chat}{\hat{C}}
\newcommand{\Dhat}{\hat{D}}
\newcommand{\Ehat}{\hat{E}}
\newcommand{\Lhat}{\hat{L}}
\newcommand{\Nhat}{\hat{N}}
\newcommand{\Ohat}{\hat{O}}
\newcommand{\Odag}{\hat{O}^{\dagger}}
\newcommand{\Shat}{\hat{S}}
\newcommand{\Uhat}{\hat{U}}
\newcommand{\Udag}{\hat{U}^{\dagger}}
\newcommand{\Xhat}{\hat{X}}
\newcommand{\Zhat}{\hat{Z}}
\newcommand{\Xdag}{\hat{X}^{\dagger}}
\newcommand{\Ydag}{\hat{Y}^{\dagger}}
\newcommand{\Zdag}{\hat{Z}^{\dagger}}
\newcommand{\Ham}{\hat{H}}
\newcommand{\bis}{{\prime \prime}}
\newcommand{\tris}{{\prime \prime \prime}}
\newcommand{\ket}[1]{\mbox{$|#1\rangle$}}
\newcommand{\bra}[1]{\mbox{$\langle#1|$}}
\newcommand{\ketbra}[2]{\mbox{$|#1\rangle \langle#2|$}}
\newcommand{\braket}[2]{\mbox{$\langle#1|#2\rangle$}}
\newcommand{\bracket}[3]{\mbox{$\langle#1|#2|#3\rangle$}}
\newcommand{\mat}[1]{\overline{\overline{#1}}}
\newcommand{\dotp}{\mbox{\boldmath $\cdot$}}
\newcommand{\tp}{\otimes}
\newcommand{\op}[2]{\mbox{$|#1\rangle\langle#2|$}}
\newcommand{\hak}[1]{\left[ #1 \right]}
\newcommand{\vin}[1]{\langle #1 \rangle}
\newcommand{\abs}[1]{\left| #1 \right|}
\newcommand{\tes}[1]{\left( #1 \right)}
\newcommand{\braces}[1]{\left\{ #1 \right\}}
\newcommand{\up}{\ket{\uparrow}}
\newcommand{\down}{\ket{\downarrow}}
\newcommand{\upx}{\ket{x_+}}
\newcommand{\downx}{\ket{x_-}}
\newcommand{\nav}{\langle \hat{n} \rangle}

\hyphenation{Teich}

\title{Two-photon interference from two blinking quantum emitters} 
\author{Klaus D. J\"{o}ns}
\email[e-mail: ]{klausj@kth.se}
\author{Katarina Stensson}
\affiliation{Department of Applied Physics, Royal Institute of Technology (KTH),\\
AlbaNova University Center, SE - 106 91 Stockholm, Sweden}
\author{Marcus Reindl}
\affiliation{Institute of Semiconductor and Solid State Physics, Johannes Kepler University Linz, 4040, Austria}
\author{Marcin Swillo}
\affiliation{Department of Applied Physics, Royal Institute of Technology (KTH),\\
AlbaNova University Center, SE - 106 91 Stockholm, Sweden}
\author{Yongheng Huo}
\affiliation{Institute for Integrative Nanosciences, IFW Dresden, 01069, Germany}
\affiliation{Institute of Semiconductor and Solid State Physics, Johannes Kepler University Linz, 4040, Austria}
\affiliation{Hefei National Laboratory for Physical Sciences at Microscale, University of Science and Technology Shanghai, 201315, China}
\author{Val Zwiller}
\affiliation{Department of Applied Physics, Royal Institute of Technology (KTH),\\
AlbaNova University Center, SE - 106 91 Stockholm, Sweden}
\author{Armando Rastelli}
\author{Rinaldo Trotta}
\email[e-mail: ]{rinaldo.trotta@jku.at}
\affiliation{Institute of Semiconductor and Solid State Physics, Johannes Kepler University Linz, 4040, Austria}
\author{Gunnar Bj\"{o}rk}
\email[e-mail: ]{gbjork@kth.se}
\affiliation{Department of Applied Physics, Royal Institute of Technology (KTH),\\
AlbaNova University Center, SE - 106 91 Stockholm, Sweden}

\date{\today}

\begin{abstract}
We investigate the effect of blinking on the two-photon interference measurement from two independent quantum emitters. We find that blinking significantly alters the statistics in the Hong-Ou-Mandel second-order intensity correlation function g$^{(2)}(\tau)$ and the outcome of two-photon interference measurements performed with independent quantum emitters. We theoretically demonstrate that the presence of blinking can be experimentally recognized by a deviation from the g$^{(2)}_{D}(0)=0.5$ value when distinguishable photons from two emitters impinge on a beam splitter. Our findings explain the significant differences between linear losses and blinking for correlation measurements between independent sources and are experimentally verified using a parametric down-conversion photon-pair source. We show that blinking imposes a mandatory cross-check measurement to correctly estimate the degree of indistinguishability of photons emitted by independent quantum emitters.

\end{abstract}

\maketitle

\section{Introduction}
Many applications of quantum optics are based on interference of indistinguishable photons. Notably, successful two-photon interference is a prerequisite for the realization of
quantum networks~\cite{Kimble:2008}, to generate N00N states for photonic quantum simulations~\cite{Loredo.Broome.ea:2017,He.Ding.ea:2017} and sensing~\cite{Bennett.Lee.ea:2016,Muller.Vural.ea:2017}, as well as linear optics quantum computation~\cite{Knill.Laflamme.ea:2001}. Since the discovery of the underlying Hong-Ou-Mandel effect~\cite{Ghosh.Hong.ea:1986,Hong.Ou.ea:1987}, extensive research has been carried out to find the most suitable sources of single and indistinguishable photons. Although parametric down-conversion pair-sources reach near-unity visibility in two-photon interference experiments, the probabilistic emission nature of the source limits its applicability. In contrast, solid-state quantum emitters, especially self-assembled semiconductor quantum dots (QDs), can emit single-photons on demand~\cite{He.He.ea:2013,Muller.Bounouar.ea:2014} and near-unity visibility for consecutively emitted photons from the same QD has been recently reported~\cite{Somaschi.Giesz.ea:2016,Ding.He.ea:2016}. However, applications in quantum information processing and quantum networks~\cite{Cirac.Zoller.ea:1997}, as well as boosting the performance of  boson-sampling machines~\cite{Wang.He.ea:2017} will require multiple single-photon sources. Therefore, there is an ongoing effort to increase the non-optimal visibilities of two-photon interference reported in experiments performed with independent solid-state quantum emitters~\cite{Lettow.Rezus.ea:2010,Flagg.Muller.ea:2010,Patel.Bennett.ea:2010,Bernien.Childress.ea:2012,Giesz.Portalupi.ea:2015,Reindl.Joens.ea:2017}. Despite the enormous progress made on the source side, the effect of blinking~\cite{Santori.Fattal.ea:2004,Frantsuzov.Kuno.ea:2008,Davanco.Hellberg.ea:2014,Efros.Nesbitt:2016}, i.e., the intermittency in the emission of single-photons from the source, on the two-photon interference has been neglected so far. Here, we theoretically show that blinking significantly changes the outcome of the two-photon interference correlation measurement. Long term blinking, since it is a memory effect, cannot be seen as a linear loss and thus changes the ratio between the coincidences measured at zero time delay and larger time delays. We demonstrate that in the presence of blinking the measured value of the second-order intensity correlation function g$^{(2)}_{D}(0)$ for distinguishable photons from independent emitters impinging on a beam splitter differs substantially from the theoretically-expected value of 0.5. This deviation is of fundamental importance to correctly estimate the two-photon interference visibility from photons emitted by independent quantum emitters and cannot be neglected.

\section{Quantum dot measurements}
\label{Sec: QD measurement}

\begin{figure}[ht]
\begin{centering}
\includegraphics[width=0.99\linewidth]{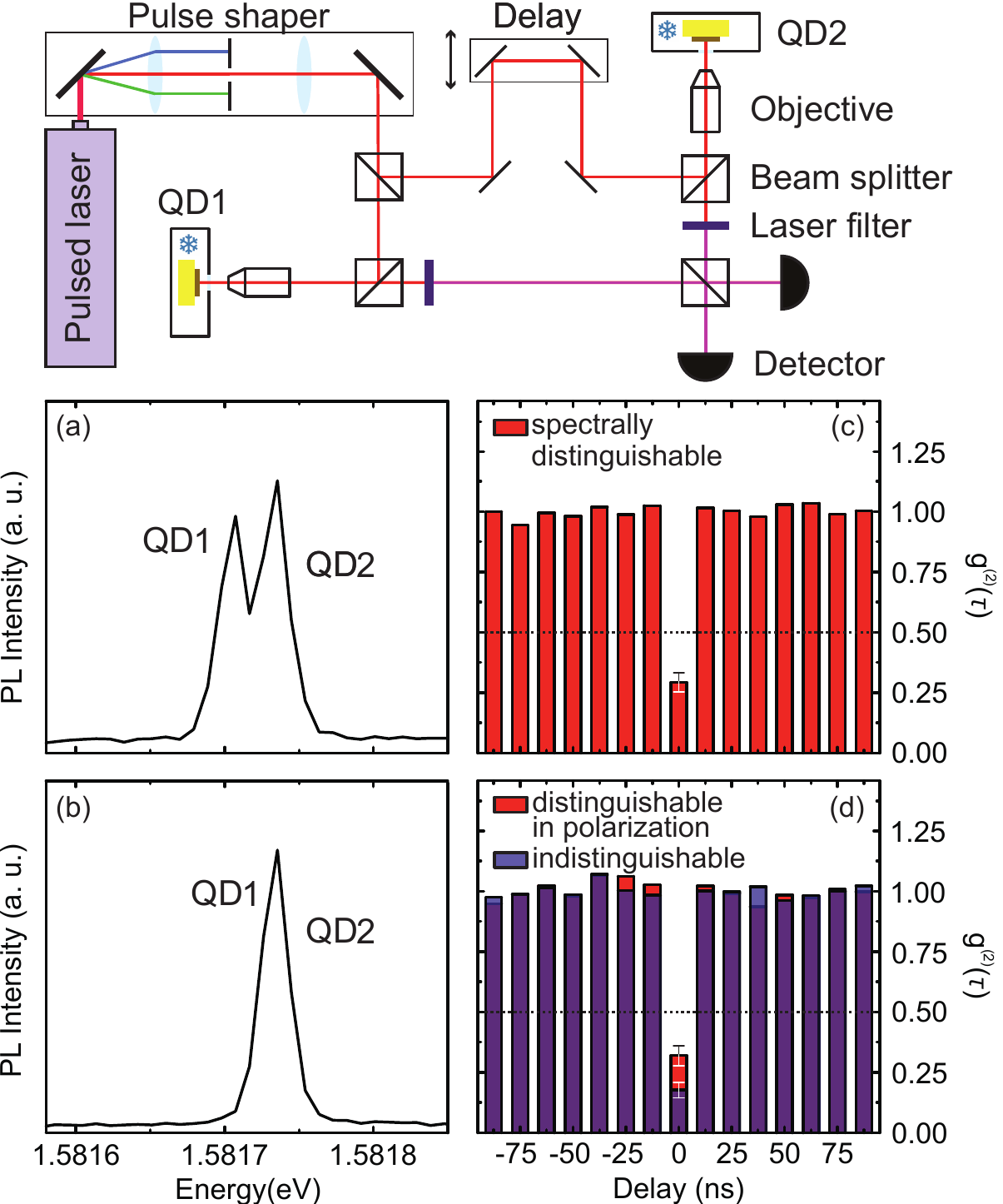}
\caption{\label{fig:fig1} Top: Schematic of the experimental setup to measure two-photon interference between two remote quantum dots. QD\,1 is mounted on a piezo-electric actuator inside the cryostat (snowflake) to allow for strain-tuning of its emission energy. (a) Photoluminescence spectrum of the neutral exciton transitions from two remote QDs. The photons stemming from these transitions do not spectrally overlap and are fully distinguishable. (b) Spectrum of the same transitions when the exciton transition of QD\,1 is strain-tuned in resonance with the exciton transition of QD\,2. (c) Normalized second-order intensity correlation measurement between spectrally distinguishable photons emitted from the transitions shown in (a). (d) Same as in (c) when the two transitions are tuned in energetic resonance. The blue data is taken when both photons have the same polarization, i.e. the photons are indistinguishable. The red data is taken when the photons have perpendicular polarization, i.e the photons are fully distinguishable. The dashed line represents the theoretically expected value of 0.5 of the center peak for fully distinguishable photons.}
\end{centering}
\end{figure}

We focus our experimental quantum dot study on symmetric GaAs/AlGaAs QDs grown via the droplet-etching method~\cite{Huo.Rastelli.ea:2013}. A detailed description of the sample structure can be found in~\cite{Huber.Reindl.ea:2017}. The QDs are excited via the phonon-assisted two-photon excitation~\cite{Glassl.Barth.ea:2013}, as discussed in~\cite{Reindl.Joens.ea:2017}. When performing two-photon interference measurements from independent QDs, we take advantage of the strain-tuning 
technique~\cite{Seidl.Kroner.ea:2006,Joens.Hafenbrak.ea:2011,Rastelli.Ding.ea:2012,Trotta.Atkinson.ea:2012} to tune the emission energy of the two transitions from the independent QDs into resonance. The effect of strain-tuning can be seen in the spectra of Fig.\,\ref{fig:fig1}\,(a)\,\&\,(b). We start with two spectrally separated neutral excitonic transitions from two QDs, where the emitted photons are fully distinguishable in energy (we note that the linewidths of the transitions from these QDs are typically an order of magnitude smaller than the spectral resolution of our spectrometer). By applying external stress to QD\,1 we can spectrally overlap both transitions (as shown in Fig.\,\ref{fig:fig1}\,(b)), making the photons partially indistinguishable. Note that equal energy and polarization represent a necessary - but not sufficient - condition for having indistinguishable photons since dephasing processes or differences in the temporal extent of the photon wavepackets make photons partially distinguishable. For simplicity we will in the following refer to the case in which polarization and energy are equal as the ``indistinguishable case". To investigate the degree of indistinguishability between the photons emitted from QD\,1 and QD\,2, we perform start-stop correlation measurements. Within our experimental conditions (low detection probability and low excitation power so that each detector event corresponds to a single impinging photon) such a start-stop experiment gives a good approximation of the second-order intensity correlation function g$^{(2)}$~\cite{Davidson.Mandel:1968}. We investigate three cases: The photons are energetically not overlapping (distinguishable case 1, shown in Fig.\,\ref{fig:fig1}\,(c) as a red bar plot), the photons are energetically overlapping and have the same polarization (indistinguishable case, shown in Fig.\,\ref{fig:fig1}\,(d) as a blue bar plot), and the photons are energetically overlapping but have perpendicular polarizations (distinguishable case 2, shown in Fig.\,\ref{fig:fig1}\,(d) as a red bar plot). During all correlation measurements we keep the average single-photon detection rate of both QD transitions equal (for the importance of this requirement see the theory section~\ref{Sec: Assumptions} in the following).
In Fig.\,\ref{fig:fig1}\,(c)\,\&\,(d) we plot the normalized coincidence counts integrated within 4\,ns time bins around every laser pulse repetition cycle. We normalize the data to the mean coincidence counts of the first 7 side peaks on each side of the zero time delay peak.
In order to estimate the degree of indistinguishability only the values of the second-order intensity correlation functions at time delay zero are relevant. Interestingly, the second-order intensity correlation measurement for the distinguishable case 1 (shown in Fig.\,\ref{fig:fig1}\,(c)) does not reach the theoretical limit of g$^{(2)}_{D}(0)=0.5$ for distinguishable photons but rather g$^{(2)}_{D}(0)=0.29 \pm 0.04$. For the indistinguishable case (blue bar plot in Fig.\,\ref{fig:fig1}\,(d)) we extract a g$^{(2)}(0)=0.18 \pm 0.03$, suggesting a very high degree of indistinguishability, whose visibility V can be calculated using 
\begin{equation}
\mathrm{V=\frac{   g_{\textit{D}}^{(2)}(0) - g_{\vphantom{\textit{D}}}^{(2)}(0)   }{  g_{\textit{D}}^{(2)}(0)} }.
\end{equation}
We would like to emphasize that assuming the theoretically-expected value of g$^{(2)}_{D}(0)=0.5$ ~\cite{Gold.Thoma.ea:2014,Thoma.Schnauber.ea:2017} would lead to a much higher visibility of two-photon interference. It is therefore extremely important to understand the reasons why g$^{(2)}_{D}(0)<0.5$ occurs in our experiments with distinguishable photons. Even though one can find similar data on a different type of quantum emitter in the literature~\cite{Bernien.Childress.ea:2012}, the deviation from the theoretical limit of g$^{(2)}_{D}(0)=0.5$ has never been discussed so far to the best of our knowledge.

To verify our experimental finding and to exclude experimental artifacts, we also perform a cross-polarized two-photon interference measurement for the same two independent QDs (distinguishable case~2). We start from the energetically overlapping case (as shown in Fig.\,\ref{fig:fig1}\,(b)) and rotate the polarization of the photons from QD\,1 perpendicular to the polarization of the photons from QD\,2, making them distinguishable again. This distinguishable case~2 is shown in Fig.\,\ref{fig:fig1}\,(d) as a red bar plot. From this measurement we extract g$^{(2)}_{D}(0)=0.32 \pm 0.04$, which is comparable to the energetically detuned case (see Fig.\,\ref{fig:fig1}\,(c)). 
In addition, we note that in the case of two-photon interference measurements performed with consecutive photons from the same quantum emitter, the cross-polarized two-photon interference measurement does reach the classical limit of 0.5. Thus, we can not only exclude any experimental error in our measurements but we can also link the effect of measuring a g$^{(2)}_{D}(0)<0.5$ to the uncorrelated photon emission between fully independent quantum emitters. In the following, we will theoretically analyze the two-photon interference measurements from independent quantum emitters and show the origin of this effect.

\section{Theory}
\label{Sec: Theory}

We assume that we have two independent QDs pumped optically by a coherent state pulse-train. Since the respective pump pulses are in coherent states, there will be no quantum correlations between the QDs, so that their respective emissions will be assumed to be uncorrelated, i.e., the joint emission state will be a tensor product of the respective QD emitted states. We will also assume that each QD emits at most one photon at a time. The emitted photons are made to interfere on a 50:50 beam splitter. (We note that it is straightforward to model other mixing ratios by assigning different overall generation/propagation/detection quantum efficiencies to the two sources). The detected photons from each of the QDs are assumed to be in a single spatio-temporal mode, but they may have different polarization states. The latter degree of freedom can be used to model any other degree of distinguishability, such as spatial, temporal, or spectral mismatch.
We will treat four cases, when the detected photons are truly indistinguishable and when the photons are fully distinguishable, under the assumption that the QDs do not blink. We will then treat the same two cases under the assumption that the QDs blink, but only for the case when the characteristic blinking frequency is much smaller than the pump pulse rate and the spontaneous emission rate of the QDs.
To calculate the coincidence probabilities we will assume that we have two detectors. One is placed at ``the first'' output port of the beam splitter and it provides a start pulse. Another is placed in ``the second'' port of the beam splitter and provides the stop pulse. The interesting information is the probability distribution of the times between successive start and stop pulses, which we calculate for all the cases discussed above.

\subsection{Indistinguishable photons, no blinking}

We shall assume that each QD (labeled as $j=1,2$) emits a state described by the density matrix
\beq
\epsilon_j^2 \ket{1}\bra{1} + (1 - \epsilon_j^2) \ket{0}\bra{0}.
\label{Eq: Emitted state}
\eeq
The parameter $\epsilon_j$ will account for all losses up until the beam splitter, such as emission into other modes, imperfect coupling of the ``interesting'' mode into the subsequent optical system, component losses, and imperfect alignment. Since all of these processes can be modeled as a linear loss, they can be lumped together into a single, overall parameter without loss of generality~\cite{Mandel.Wolf:1995}.

When two such states impinge on a 50:50 beam splitter described by the unitary 2 x 2 matrix with $U_{11} = U_{22} = U_{12} = -U _{21} = 1/\sqrt{2}$, the ensuing output state $\hat{\rho}$ becomes
\begin{eqnarray}
\hat{\rho} & = & \frac{\epsilon_1^2 \epsilon_2^2}{2}(\ket{2,0} - \ket{0,2}) \otimes \textrm{H.C.} \nonumber \\
& & + \frac{\epsilon_1^2 (1 - \epsilon_2^2)}{2}(\ket{1,0} + \ket{0,1}) \otimes \textrm{H.C.} \nonumber \\
& & + \frac{\epsilon_2^2 (1 - \epsilon_1^2)}{2}(\ket{1,0} - \ket{0,1}) \otimes \textrm{H.C.} \nonumber \\
& & + (1 - \epsilon_1^2) (1 - \epsilon_2^2)\ket{0,0} \otimes \textrm{H.C.},
\label{Eq: Beam splitter state}
\end{eqnarray}
where H.C. denotes the Hermitian conjugate of the factor to the left and, e.g. $|2,0\rangle$ denotes a product state of two photons in the detected mode exiting the ``first" beam splitter port and no photon exiting the second port.

Suppose that at least one photon is detected at the first port at time $t=0$. The state in the second mode then instantly collapses onto the state $\ket{0}\bra{0}$. This means that in this case there can be no coincidence between the two detectors detected at the same time (meaning, in practice, within the spontaneous emission time of the QDs). We thus conclude that the probability $p(0)$ of getting a stop pulse at $t=0$ is $p(0)=0$.

However, one pump pulse later, at $t=\tau$, there is anew a state as in Eq. (\ref{Eq: Emitted state}) emitted from each QD, which after the beam splitter will have the form of Eq. (\ref{Eq: Beam splitter state}). Since this state is uncorrelated to the state at $t=0$, the probability $p$ of detecting at least one photon at the second beam splitter output port at $t=\tau$ is
\begin{eqnarray}
p(\tau) & &= \eta_2^2 \left [ \frac{\epsilon_1^2 \epsilon_2^2 (2-\eta_2^2)}{2} + \frac{\epsilon_1^2 (1 - \epsilon_2^2)}{2} + \frac{\epsilon_2^2 (1 - \epsilon_1^2)}{2}\right ] \nonumber \\
& & = \eta_2^2 \frac{\epsilon_1^2 + \epsilon_2^2 - \eta_2^2\epsilon_1^2 \epsilon_2^2}{2},
\end{eqnarray}
where $\eta_2^2$ ($\eta_1^2$) is the detection efficiency at the second (first) beam splitter output port. Should the second detector not detect a photon, which happens with probability $1-p(\tau)$, it has a new chance at time $t= 2 \tau$. The probability of detecting a photon at the second output port at that time is $p(2 \tau)=p(\tau)[1-p(\tau)]$. (Note that if the photon is detected at time $\tau$, photon counting restarts, so that we must consider the conditional probability). Likewise, the probability of detecting a photon at time $t = m \tau$ will be $p(m \tau)=p(\tau) [1-p(\tau)]^{m-1}$, $m= 1, 2, \ldots$.

\noindent In the limit when $\epsilon_1 = \epsilon_2 = \epsilon \ll 1$, the probability of a stop pulse at $t=\tau$ simplifies to $p(\tau)=\eta_2^2 \epsilon^2$.

\subsection{Distinguishable photons, no blinking}
\label{Sec: Distinguishable no blinking}

In this case we shall assume that the two QDs emit states described by
\beq
\epsilon_1^2 \ket{V}\bra{V} + (1 - \epsilon_1^2) \ket{0}\bra{0},
\label{Eq: Vertical emitted state}
\eeq
and
\beq
\epsilon_2^2 \ket{H}\bra{H} + (1 - \epsilon_2^2) \ket{0}\bra{0},
\label{Eq: Horizonal emitted state}
\eeq
respectively. Here, e.g., $\ket{V}$ denotes one photon in vertical polarization. This state is orthogonal to the state $\ket{H}$ meaning that they are single shot, 100 \% distinguishable and thus they will not interfere. Below we shall also use the notation $\ket{VH}$ that denotes one vertically and one horizontally polarized photon in one spatio-temporal mode. By assuming that the two states above impinge on a 50:50 beam splitter one arrives at the state $\hat{\rho}'$ given by
\begin{eqnarray}
\hat{\rho}' & = & \frac{\epsilon_1^2 \epsilon_2^2}{4}(\ket{VH,0} - \ket{V,H} + \ket{H,V} - \ket{0,VH}) \otimes \textrm{H.C.} \nonumber \\
& & + \frac{\epsilon_1^2 (1 - \epsilon_2^2)}{2}(\ket{V,0} + \ket{0,V}) \otimes \textrm{H.C.} \nonumber \\
& & + \frac{\epsilon_2^2 (1 - \epsilon_1^2)}{2}(\ket{H,0} - \ket{0,H}) \otimes \textrm{H.C.} \nonumber \\
& & + \frac{(1 - \epsilon_1^2) (1 - \epsilon_2^2)}{2}\ket{0,0} \otimes \textrm{H.C.}.
\end{eqnarray}
\begin{widetext}
Suppose now that at least one photon, irrespective of polarization, is detected at the first beam splitter output at time $t=0$. The state at the second beam splitter output port then collapses into the state
\beq
\frac{1}{N}\left [ \frac{\epsilon_1^2 \epsilon_2^2}{4}\left (\ket{V}\bra{V}  + \ket{H}\bra{H} - \left [2+\eta_1^2 \right ] \ket{0}\bra{0} \right) +  \left( \frac{\epsilon_1^2 +\epsilon_2^2 }{2}\right) \ket{0}\bra{0}\right ],
\eeq
where the state normalization factor is $N=(\epsilon_1^2 +\epsilon_2^2)/2 -\eta_1^2 \epsilon_1^2 \epsilon_2^2/4$.
\end{widetext}
Thus, the probability $p_D(0)$ of detecting one photon also at the second beam splitter port at $t=0$ becomes
\beq
p_D(0) =  \frac{2\eta_2^2 \epsilon_1^2 \epsilon_2^2}{2 (\epsilon_1^2 + \epsilon_2^2) - \eta_1^2 \epsilon_1^2 \epsilon_2^2}.
\eeq
At time $t=\tau$, uncorrelated states are emitted, so given the situation that no photon was detected at the second beam splitter port at $t=0$ (that happens with probability $1-p_D(0)$), the probability of detecting the stop pulse at $ t=\tau $ becomes
\begin{eqnarray}
p_D(\tau)& &=  \eta_2^2 [1-p_D(0)] \frac{2(\epsilon_1^2 + \epsilon_2^2) - \eta_2^2\epsilon_1^2\epsilon_2^2}{4} \nonumber \\
& & = \eta_2^2 [1-p_D(0)]p_D,
\end{eqnarray}

where $p_D = \frac{2(\epsilon_1^2   + \epsilon_2^2  )-\eta_2^2  \epsilon_1^2  \epsilon_2^2}{4}$.
In the same manner as for indistinguishable photons, the probability of detecting the first stop pulse at $t=2 \tau$ becomes $p_D(2\tau)=\eta_2^2 [1-p_D(0)][1-p_D]p_D$ and the probability that the stop pulse comes at time $t=m \tau$ is $p_D(m \tau)=\eta_2^2 [1-p_D(0)](1-p_D)^{m-1}p_D$.

In the limit when $\epsilon_1 = \epsilon_2 = \epsilon \ll 1$, the probability of a stop pulse at $t=0$ becomes $p_D(0) = \eta_2^2 \epsilon^2/2$ and the probability of a stop pulse at $t=\tau$ simplifies to $p_D(\tau)=\eta_2^2 \epsilon^2$. Hence the ratio $p_D(0)/p_D(\tau) = 1/2$, which is the theoretically-expected value g$_D^{(2)}(0)$ we have mentioned in section~\ref{Sec: QD measurement}.

\subsection{Indistinguishable photons, blinking}

Here we introduce the further complication that the QDs may blink, that is, they randomly (in time) enter a ``dark state'', i.e., a state where they do not emit any photons at all. If the typical times between the transitions from a bright to a dark state, or vice versa, are very long compared to the other time scales of the problem, then we can neglect the situations where one QD makes such a transition right after the start pulse is detected. Hence we will only consider the four possibilities that none of the QDs are dark when the start pulse is detected until a stop pulse is detected, that only one of the two QDs is dark (but that the other remains in an emitting state for the entire duration between a start and a stop pulse), and that both of them are dark. If we make the reasonable assumption that the two QDs blink independently, and assume that the probabilities of QD $j=1,2$ to be in the on-state to be $\pi_j$, then the probability of detecting a coincidence at time $t=0$ is still zero since a coincident detection cannot happen for indistinguishable photons irrespective if both QDs are emitting or if only one of them is emitting. Thus, the corresponding probability $p'(0)=0$, where the prime indicates that the QDs are assumed to be blinking.

Given that we detected at least one photon at the first beam splitter port (giving the start signal), the state at the detectors at $t=\tau$ is independent of the ``start event''. To detect a photon at the second output port given that there was a start pulse (eliminating the possibility that both QDs were in there dark state) requires that either only the first, only the second, or both the QDs are emitting. The conditional probabilities for this are $\pi_1(1-\pi_2)/(\pi_1 + \pi_2 - \pi_1\pi_2)$, $\pi_2(1-\pi_1)/(\pi_1 + \pi_2 - \pi_1\pi_2)$ and $\pi_1 \pi_2/(\pi_1 + \pi_2 - \pi_1\pi_2)$, respectively. In the first two cases the probability for finding an emitted photon at the second port is $\epsilon_i^2/2$. Hence, the probability of detecting the stop pulse at time $t=\tau$ becomes
\begin{eqnarray}
p'(\tau) & = &\eta_2^2 \left [ \frac{\pi_1(1-\pi_2)}{\pi_1 + \pi_2 - \pi_1\pi_2} \frac{\epsilon_1^2}{2} + \frac{\pi_2(1-\pi_1)}{\pi_1 + \pi_2 - \pi_1\pi_2} \frac{\epsilon_2^2}{2} \right. \nonumber \\
& &\left. + \frac{\pi_1\pi_2}{\pi_1 + \pi_2 - \pi_1\pi_2} \frac{\epsilon_1^2 + \epsilon_2^2-\eta_2^2 \epsilon_1^2\epsilon_2^2}{2}\right ].
\end{eqnarray}
The probability of detecting the stop pulse at $t=2\tau$ becomes
\begin{eqnarray}
p'(2\tau) & = & \eta_2^2 \left [ \frac{\pi_1(1-\pi_2)}{\pi_1 + \pi_2 - \pi_1\pi_2} \left ( 1-\frac{\eta_2^2\epsilon_1^2}{2} \right ) \frac{\epsilon_1^2}{2} \right. \nonumber \\
 & & +\frac{\pi_2(1-\pi_1)}{\pi_1 + \pi_2 - \pi_1\pi_2} \left ( 1-\frac{\eta_2^2\epsilon_2^2}{2} \right ) \frac{\epsilon_2^2}{2} \nonumber \\
 & & +\frac{\pi_1\pi_2}{\pi_1 + \pi_2 - \pi_1\pi_2} \left ( 1-\eta_2^2\frac{\epsilon_1^2 + \epsilon_2^2-\eta_2^2\epsilon_1^2\epsilon_2^2}{2} \right ) \nonumber \\
 & & \left. \times \frac{\epsilon_1^2 + \epsilon_2^2-\eta_2^2\epsilon_1^2\epsilon_2^2}{2} \right ].
 \end{eqnarray}
For $t= m \tau$ the probability becomes
\begin{eqnarray}
p'(m \tau) & = & \eta_2^2 \left [  \frac{\pi_1(1-\pi_2)}{\pi_1 + \pi_2 - \pi_1\pi_2} \left ( 1-\frac{\eta_2^2\epsilon_1^2}{2} \right )^{m-1} \frac{\epsilon_1^2}{2} \right. \nonumber \\
 & & +\frac{\pi_2(1-\pi_1)}{\pi_1 + \pi_2 - \pi_1\pi_2} \left ( 1-\frac{\eta_2^2\epsilon_2^2}{2} \right )^{m-1} \frac{\epsilon_2^2}{2}  \nonumber \\
  & & + \frac{\pi_1\pi_2}{\pi_1 + \pi_2 - \pi_1\pi_2} \left ( 1-\eta_2^2\frac{\epsilon_1^2 + \epsilon_2^2-\eta_2^2\epsilon_1^2\epsilon_2^2}{2} \right )^{m-1} \nonumber \\
 & &\left.  \times \frac{\epsilon_1^2 + \epsilon_2^2-\eta_2^2\epsilon_1^2\epsilon_2^2}{2} \right ] .
 \end{eqnarray}

In the limit $\epsilon_1 = \epsilon_2 = \epsilon \ll 1$, and assuming that $\pi_1 = \pi_2 = 1/2$ (i.e. that the QDs spent half of the time in a bright and half in a dark state) the probability of a stop pulse at $t=\tau$ becomes $p'(\tau) = 2 \eta_2^2 \epsilon^2/3$. If instead we assume that $\pi_1 = \pi_2 = \pi \ll 1$ (i.e. the QDs are most of the time in a dark state), then we arrive at the result $p'(\tau) =\eta_2^2 \epsilon^2/2$. The reason the results are independent of the on-state probability $\pi$ is that in order to have a start pulse, at least one QD must be in the on-state. Thus, when looking for a stop pulse, we know already that at least one of the QDs is emitting, eliminating the unconditional probability $\propto \pi$ that this is the case at any time.

We also note that for perfectly indistinguishable, emitted photons, the probability of detecting a coincidence at $t=0$ remains zero, no matter if the two emitters are blinking or not. Both probabilities $p(0)$ and $p'(0)$ vanish. This is an obvious result, since both emitters emit single photons (so that two photons from the same dot cannot be detected at the two output ports) and indistinguishability implies that photons emitted by the two QDs exit the same port of the beam splitter. It is therefore not possible to distinguish between these two possibilities (blinking/non-blinking) based on a comparison between the ratios $p(0)/p(\tau)$ and $p'(0)/p'(\tau)$ since both are ideally zero.

\subsection{Distinguishable photons, blinking}
\label{Sec: Distinguishable photons, blinking}
In this case we can compute the probability for a coincidence at $t=0$ directly from the result in Sec. \ref{Sec: Distinguishable no blinking}. We first note that in order to get a start pulse at the detector at the first beam splitter port either one, the other, or both QDs must be in their emitting states. The conditional probabilities for this are $\pi_1(1-\pi_2)/(\pi_1 + \pi_2 - \pi_1\pi_2)$, $\pi_2(1-\pi_1)/(\pi_1 + \pi_2 - \pi_1\pi_2)$ and $\pi_1 \pi_2/(\pi_1 + \pi_2 - \pi_1\pi_2)$. Noting that in order to get a stop pulse at $t=0$ to be a possibility, neither of the QDs can be in their dark state. The probability for this, given that we had a start pulse, is
\beq
p'_D(0) = \eta_2^2 \frac{\pi_1 \pi_2}{\pi_1 + \pi_2 - \pi_1 \pi_2} \frac{2\epsilon_1^2 \epsilon_2^2}{2 (\epsilon_1^2 + \epsilon_2^2) - \eta_1^2 \epsilon_1^2 \epsilon_2^2}.
\label{Eq: distinguish blinking}
\eeq

The probability of detecting at least one photon at the second beam splitter port at time $t = \tau$, given that one photon was detected at the first port but none at the second at $t=0$ is given by
\begin{eqnarray}
p'_D(\tau)& = & \eta_2^2 \left [ \frac{\pi_1 (1-\pi_2)}{\pi_1 + \pi_2 - \pi_1 \pi_2} \frac{\epsilon_1^2}{2} + \frac{\pi_2 (1-\pi_1)}{\pi_1 + \pi_2 - \pi_1 \pi_2} \frac{\epsilon_2^2}{2} \right. \nonumber \\
& & + \frac{\pi_1 \pi_2}{\pi_1 + \pi_2 - \pi_1 \pi_2}  \left (1-\frac{2\eta_2^2\epsilon_1^2 \epsilon_2^2}{2 (\epsilon_1^2 + \epsilon_2^2) - \eta_1^2 \epsilon_1^2 \epsilon_2^2} \right ) \nonumber \\
& & \left. \times  \frac{2(\epsilon_1^2 + \epsilon_2^2) - \eta_2^2\epsilon_1^2\epsilon_2^2}{4} \right ],
\end{eqnarray}
where we have used the fact that the emission from subsequent pump pulses is uncorrelated, that if we detect a start pulse, this rules out the possibility of having both QDs in their respective dark states, and if only one QD is in its dark state, we cannot get a coincidence at $t=0$.

The probability of getting a stop pulse at $t= m \tau$, $m=1,2, \ldots$ becomes, in this case
\begin{eqnarray}
p'_D(m  \tau)& = & \eta_2^2 \left [ \frac{\pi_1 (1-\pi_2)}{\pi_1 + \pi_2 - \pi_1 \pi_2} \left ( 1-\frac{\eta_2^2\epsilon_1^2}{2} \right )^{m-1}\frac{\epsilon_1^2}{2} \right. \nonumber \\
& & + \frac{\pi_2 (1-\pi_1)}{\pi_1 + \pi_2 - \pi_1 \pi_2} \left ( 1-\frac{\eta_2^2\epsilon_2^2}{2} \right )^{m-1}\frac{\epsilon_2^2}{2} \nonumber \\
& & + \frac{\pi_1 \pi_2}{\pi_1 + \pi_2 - \pi_1 \pi_2}  \left (1-\frac{2\eta_2^2\epsilon_1^2 \epsilon_2^2}{2 (\epsilon_1^2 + \epsilon_2^2) - \eta_1^2\epsilon_1^2 \epsilon_2^2} \right ) \nonumber \\
& & \times \left ( 1- \eta_2^2\frac{\epsilon_1^2 + \epsilon_2^2 - \eta_2^2\epsilon_1^2\epsilon_2^2}{2} \right )^{m-1}  \nonumber \\
& & \left. \times \frac{2(\epsilon_1^2 + \epsilon_2^2) - \eta_2^2\epsilon_1^2\epsilon_2^2}{4} \right ].
\end{eqnarray}

\begin{figure}[th]
\begin{centering}
\includegraphics[width=0.99\linewidth]{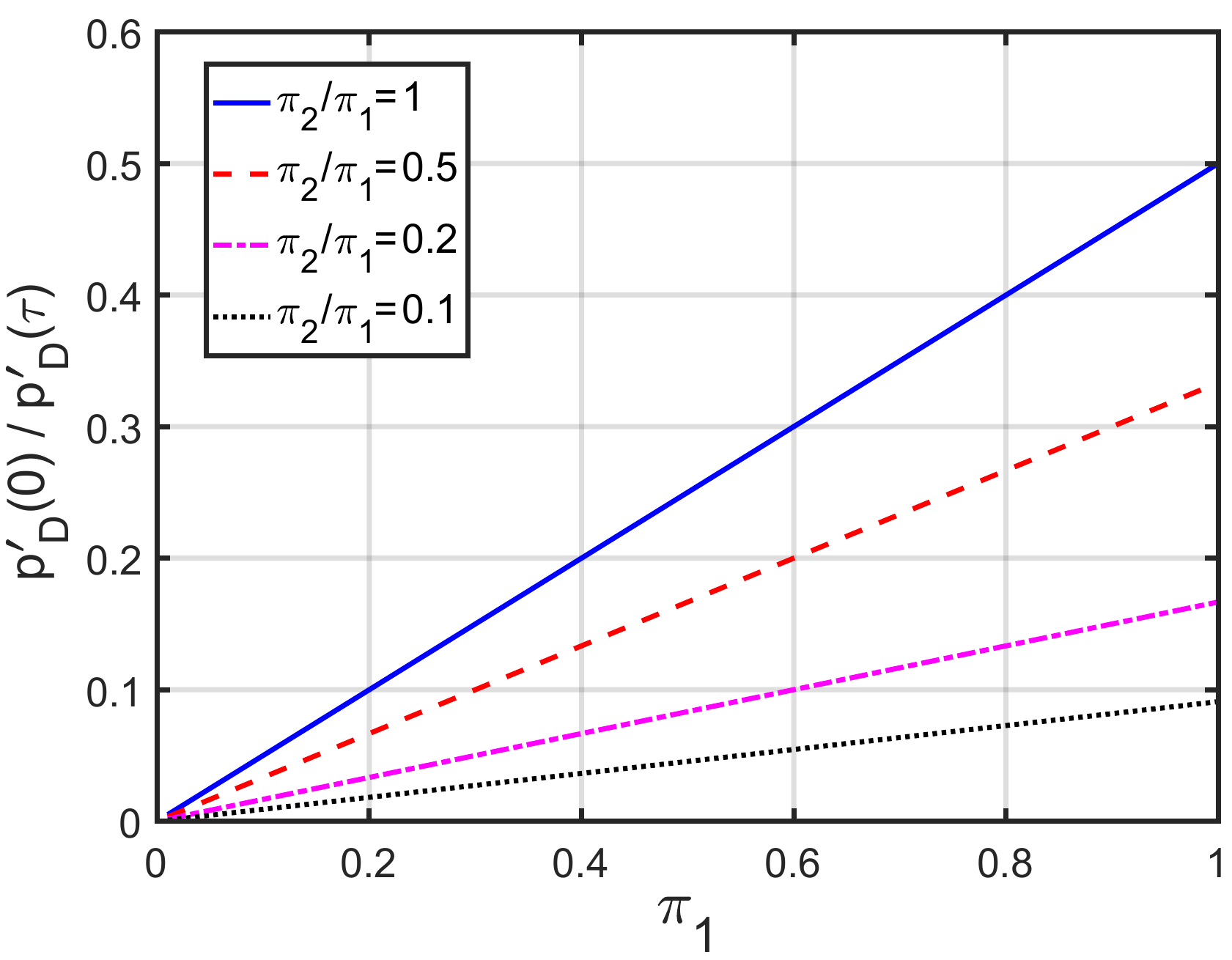}
\caption{\label{fig:fig2} Calculated $p'_D(0)/p'_D(\tau)$ ratio as a function $\pi_1$, which is the probability that QD~$j=1$ is in the on-state. The different curves represent different on-state ratios ($\pi_2/\pi_1$) between the two QDs $j=1,2$. It is assumed that $\epsilon_1= \epsilon_2 \ll 1$ and $\eta_1=\eta_2=1$.}
 \end{centering}
\end{figure}

In the limit $\epsilon_1 = \epsilon_2 = \epsilon \ll 1$, and assuming that $\pi_1 = \pi_2 = 1/2$ the probability of a stop pulse at $t=0$ becomes $p'_D(0) = \eta_2^2 \epsilon^2/6$ and getting a stop pulse at $t = \tau$ is $p'_D(\tau) = 2\eta_2^2 \epsilon^2/3$. Hence the ratio between these two probabilities is $1/4$, smaller than in the non-blinking case and thus a factor of 1/2 smaller than the normalized second order correlation function $g_D^{(2)}(0)$ one would have expected from distinguishable, non-blinking photons. If instead we assume that $\pi_1 = \pi_2 = \pi \ll 1$, then we arrive at the probability $p'_D(0) = \pi \eta_2^2 \epsilon^2/4$ to get a stop pulse at $t=0$ and the probability $p'_D(\tau) = \eta_2^2 \epsilon^2/2$ to get the stop pulse at $t=\tau$. In this case the ratio between the probabilities is $\pi/2$. The latter can be very small if the on-state probability $\pi$ is small. In Fig.\,\ref{fig:fig2} we plot the $p'_D(0)/p'_D(\tau)$ ratio as a function $\pi_1$ for four different on-state ratios ($\pi_2/\pi_1$). In the presence of blinking the $p'_D(0)/p'_D(\tau)$ ratio is clearly smaller than 0.5, leading to g$^{(2)}_{D}(0)<0.5$, as observed in the quantum dot experiments. One sees that for blinking emitters, the ratio $p'_D(0)/p'_D(\tau)$ can be arbitrarily small. 

For non-blinking emitters a ratio below 1/2 would indicate that the emitted photons were partly indistinguishable, or that the efficiency (e.g. setup losses, collection efficiency,...) of the two sources are not equal.
It is important to point out the latter explanation cannot be invoked to explain the deviation from 0.5 observed in the experimental data of Fig.~\ref{fig:fig1}. In fact, this hypothesis would imply more than a factor 2 difference in the efficiency of the two QDs, something that we have experimentally ruled out (see section~\ref{Sec: QD measurement}). However, for blinking emitters a deviation from the theoretical value of 0.5 can be observed even when the efficiencies are kept the same. In this case, the proper way of assessing the indistinguishability of the emitted photons is to make the emitters perfectly distinguishable (e.g., by transforming them into mutually orthogonal polarization states) and measuring the $p'_D(0)/p'_D(\tau)$ ratio. Subsequently one makes the photons as indistinguishable as possible and re-measures this ratio. \textit{Only the comparison between these two ratios will quantify the indistinguishability of the emitted photons.}

\subsection{Comments}
We note that for inefficient QDs (time spent in dark state much larger than time spent in bright state, i.e. $\pi<<1$) the probability ratio between getting a stop pulse at $t=0$ and getting a stop pulse at $t=\tau$ goes to zero. The reason is that, in order to detect a stop pulse at $\tau=0$ (for distinguishable photons), both QDs need to be in their on-state. 
 Whereas, to detect a stop pulse at $t=\tau$ it suffices that at least one QD is in its on-state. For small values of $\pi$, the probability ratio between these two cases is roughly $\pi$. The implication will be further explained in the following section.

\section{Implications and experimental verification}
\label{Sec: Experimental}
Let us redo the derivation of $p_D'(\tau)$ in section \ref{Sec: Distinguishable photons, blinking}, but now assume a coincidence type of measurement rather than a start-stop measurement (the full derivation is given in the appendix). The origin of the observed behavior now appears much more clearly. The probability $P_D'(\tau)$ of getting a coincidence at time delay $\tau$, when blinking is present, is in this case given by

\begin{equation}
P_D'(\tau) = \frac{r^2 \eta_1^2 \eta_2^2}{2} \left [ \pi_1 \epsilon_1^4 + \pi_2 \epsilon_2^4 + 2 \pi_1 \pi_2 \epsilon_1^2\epsilon_2^2  \right],
\label{Eq: Distinguishable appendix}
\end{equation}

where $r$ is a constant relating to the emission probability. The last term corresponds to coincidences of photons from two separate emitters. This term scales equally with the blinking on-state probabilities $\pi_i$ and the total quantum efficiencies $\epsilon_i^2$. In this contribution to $P_D'(\tau)$, you can not distinguish photons ``lost" through blinking from linear loss. However, the first two terms, corresponding to coincidences of two photons emitted from the same emitter separated by a time delay $\tau$, scales linearly with blinking on state probability $\pi_i$ and quadratically with total quantum efficiency $\epsilon_i^2$. This contribution raises $P_D'(\tau)$ when the non-unity efficiency is due to blinking rather than due to linear loss, and thereby it lowers the ratio $P_D'(0)/P_D'(\tau)$, when blinking is present, since the contribution from these two terms is of course zero at zero time delay (the sources emit single photons). This lets us conclude that the origin of the deviation towards zero of $g^{(2)}(0)/g^{(2)}(\tau)$ when blinking is present lies in the inherent difference between blinking and other types of (linear) losses exhibited by the emitted photons. 

This result should be obvious when looking more closely at the following example situation: Consider only photons emitted from quantum emitter no. 1 ($\pi_2 = 0$) and assume $\pi_1=1/2$. Now, either two consecutive photons pass through the setup (when the emitter is in the on-state) or none do (when in the off-state). This would result in half the number of coincidences, at a time delay $\tau$, compared to when no blinking is present. In contrast, assuming a unity on-state probability $\pi_1=1$ but adding a filter with transmission T=1/2 results in 1/4 the number of coincidences compared to when no blinking or filtering is present, since each photon has a 1/2 probability of getting lost. 

To further clarify the difference between blinking and linear loss we performed a simple experiment, using an inherently non-blinking parametric down conversion source providing indistinguishable single photons (two-photon interference visibility V$=0.929 \pm 0.002$ (raw data)), a chopper with variable duty cycle corresponding to blinking on-state probabilities $\pi_i$, and filters with varying transmittance to vary the overall quantum efficiency $\epsilon_i^2$.

\subsection{Measurement Setup}

\begin{figure}[th]
\begin{centering}
\includegraphics[width=0.48\textwidth]{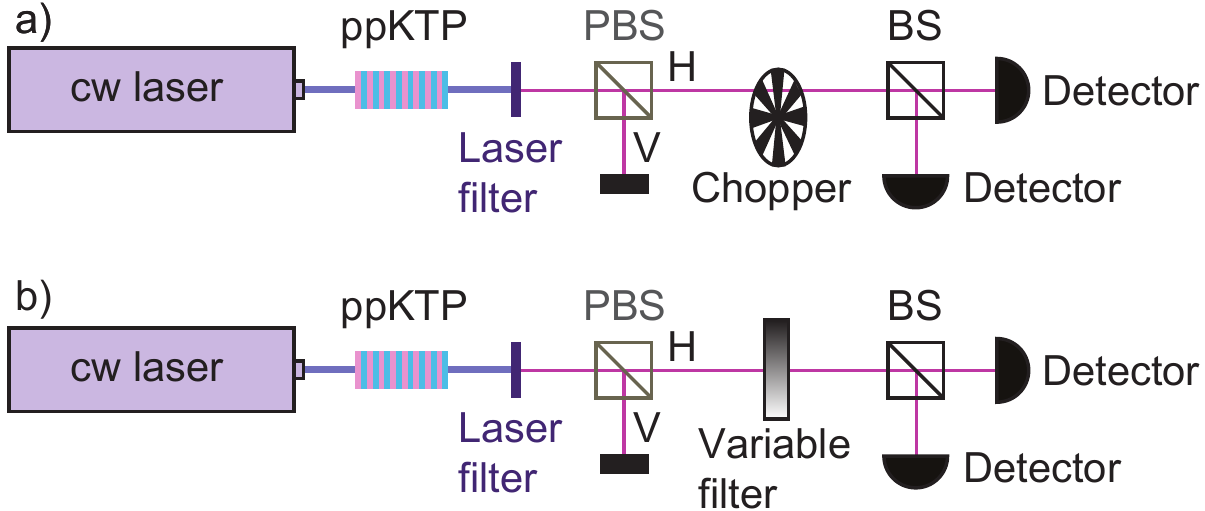}
\caption{\label{fig:setup} Schematic of the parametric down conversion setup for investigating the difference of linear losses and blinking. Either a) a chopper is inserted to mimick blinking or b) a variable linear loss filter.}
 \end{centering}
\end{figure}

A schematic of the measurement setup is depicted in Fig.\,\ref{fig:setup}. A periodically poled potassium titanyl phosphate (ppKTP) crystal is pumped with a cw laser of wavelength 405 nm. Photon pairs are spontaneously generated at 810\,nm wavelength and of perpendicular polarization. The pair is split on a polarizing beam splitter and one output is immediately blocked, while the other is let to impinge on an ordinary 50:50 beam splitter. The outputs from the beam splitter are coupled into single mode fibers, and detected by avalanche photo diodes. The quantum efficiencies of the detectors are around 50\,\% at 810\,nm.

Either a chopper (Fig.\,\ref{fig:setup}\,a) of varying duty cycles or filters with varying transmittance (Fig.\,\ref{fig:setup}\,b) can be inserted in the stream of single photons. The chopper frequency is chosen to be around 200\,Hz, corresponding to a blinking time scale of milliseconds. The duty cycle of the choppers can be varied in the interval 0-50\,\% transmission.

\subsection{Effect of Blinking VS linear loss}
\begin{figure}[th]
\begin{centering}
\includegraphics[width=0.49\textwidth]{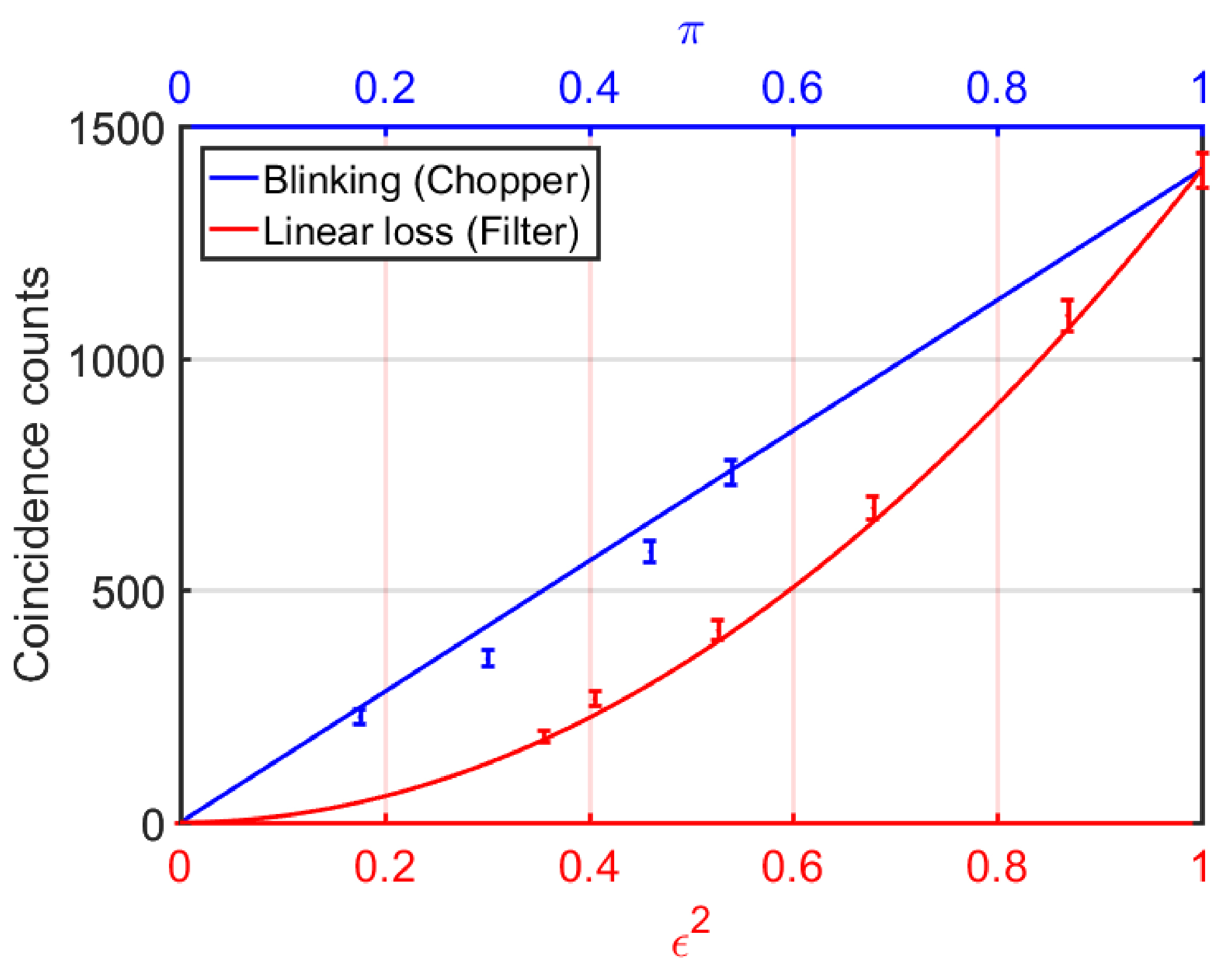}
\caption{\label{fig:blinkingvslinear} 
Effect of blinking versus linear loss. The data points are measured coincidences at time delay $\tau=300$\,ns, between photons entering the beam splitter through the same input port, as a function of induced blinking on-state probability $\pi$ (blue) and filter transmission $\epsilon^2$ (linear loss, red). Error bars account for statistical errors. Solid lines represent Eq.~\eqref{Eq: Distinguishable appendix}, setting $\pi_2=0$.}
\end{centering}
\end{figure}

The coincidence counts at time delay $\tau=300$\,ns between the two detectors were measured as a function of chopper duty cycle $\pi$ and filter transmission $\epsilon^2$. The results are presented in Fig.\,\ref{fig:blinkingvslinear}. Solid lines correspond to Eq. \eqref{Eq: Distinguishable appendix} setting $\pi_2=0$, and the pre-factor, a combination of emission rate and detector efficiency, adjusted to fit the common end points of the filter and chopper data point sequences.

The data confirms the theoretical findings that this contribution to $p_D'(\tau)$ scales linearly with the blinking on-state probability $\pi$ (chopper duty cycle) but quadratic with the linear loss $\epsilon^2$ (filter transmission). We can clearly observe that blinking and linear loss affect the correlation measurement in different ways, with the blinking resulting in proportionally higher side peaks and thus lowering the ratio $p'_D(0)/p'_D(\tau)$.

\section{Two-photon interference from a single blinking quantum dot}
We would like to give a final remark on the single quantum dot HOM-type experiments where consecutive photons from the same quantum dot are interfered in an unbalanced Mach-Zehnder interferometer to determine the photon indistinguishability. For a single emitter three situations can occur: 1) The quantum emitter is in the on-state  both at the start and the stop signal. This is described by the standard theory. 2) The quantum emitter is in the off-state. This gives no contribution to the second-order intensity correlation function at all. 3) The quantum emitter is in the on-state, giving a start signal, but switches in the off-state before we get a stop signal. The stop signal will only come as the quantum emitter returns in the on-state again. In general, this condition can be treated as for the case of blinking quantum dots we discussed above. However, this is an extremely rare event, as the blinking is typically very slow compared to the spontaneous emission decay time and pump repetition period. Thus in practice it will hardly make any contribution to the second order intensity correlation function. 
In experiments with two quantum emitters, the relevant parameter is the probability that the quantum emitter is in its on-state, not the probability that it made a transition between the start and the stop pulse. Thus the effect of blinking is significant for two blinking quantum emitter but insignificant for a single blinking emitter.

\section{Conclusion}
In our study we have shown that the second-order intensity correlation measurements between distinguishable independent quantum emitters can go below the theoretically-expected value of g$^{(2)}_{D}(0)=0.5$. We attribute this effect to the inherent blinking of the quantum emitters, which cannot be treated as linear losses. Using a parametric down conversion pair source, we experimentally verified the differences between blinking and linear losses on the second order intensity correlation function.
As the blinking behavior of quantum emitters is often unknown, it is mandatory to measure the second-order intensity correlation function for distinguishable photons impinging on a beam splitter to correctly estimate the degree of indistinguishability of photons from independent emitters.

\begin{acknowledgments}
This work was supported by the Swedish Research Council (VR) through its support of the Linn\ae us Excellence Center ADOPT and contract No. 621-2014-5410, and the ERC Starting Grant No. 679183 (SPQRel). K.D.J. acknowledges funding from the Marie Sk\l{}odowska Individual Fellowship under REA grant agreement No. 661416 (SiPhoN). K.D.J. and R.T. acknowledge the COST Action MP1403, supported by COST (European Cooperation in Science and Technology). K.D.J. would like to thank Mete Atat\"ure for fruitful discussions.
\end{acknowledgments}

\appendix*
\section{Simulating blinking with a parametric down-conversion source - Theory}
\label{Sec: Assumptions}

To experimentally test the influence of blinking under controlled conditions, we use a ppKTP crystal to generate wavelength degenerate photon pairs. The emitted photons have orthogonal polarization, but after separating the photons with a polarization beam splitter we can either rotate the polarization of one of the photons by 90 degrees, or let it stay in the orthogonal, and therefore fully distinguishable, polarization. In each of the polarization ``arms'' we can either block the beam, or insert a neutral density filter. The emitted photons are subsequently made to interfere on a 50:50 beam splitter. After the splitter the beams are focused onto photo detectors whose count rates, and coincidence count rate are recorded.

In the following, the situation when the two photons are cross-polarized, and thus fully distinguishable will be analyzed.

We shall assume that the photon pair source emits a state described by
\beq
r \ket{H,V}\bra{H,V} + (1 - r) \ket{0}\bra{0}.
\label{Eq: Appendix Emitted state}
\eeq
The parameter $r$ will account for the fact that the photon pair production is a spontaneous process, and to keep the production of four or more photons at a minimum, the pump intensity is deliberately chosen so that $r$ is below the one percent level.

The produced photons are subsequently spatially separated by a polarizing beam splitter, and each beam then suffer linear losses that can be increased by introducing neutral density filters in each arm. The total linear losses in each arm will be denoted $1-\epsilon_j^2$, $j=1,2$, where the index 1 (2) denotes the arm of the horizontally (vertically) polarized photon. The state after the attenuation will be

\begin{eqnarray}
\hat{\rho}_i & = & r \epsilon_1^2 \epsilon_2^2 \ket{H,V}\bra{H,V} + r \epsilon_1^2 (1-\epsilon_2^2)\ket{H,0}\bra{H,0}\nonumber \\
& & + r \epsilon_2^2 (1 - \epsilon_1^2)\ket{0,V}\bra{0,V} \nonumber \\
& & +(1 - r + r(1 - \epsilon_1^2)(1 - \epsilon_2^2))\ket{0,0}\bra{0,0}.
\label{
state}
\end{eqnarray}

When such a state impinge on the two input ports of a 50:50 beam splitter, the ensuing output state $\hat{\rho}_o$ becomes
\begin{eqnarray}
\hat{\rho}_o & = & \frac{r \epsilon_1^2 \epsilon_2^2}{4}\left (\ket{HV,0}\bra{HV,0} + \ket{H,V}\bra{H,V} \right. \nonumber \\
& &\left. + \ket{V,H}\bra{V,H} + \ket{0,HV}\bra{0,HV}\right )\nonumber \\
& & + \frac{r \epsilon_1^2 (1 - \epsilon_2^2)}{2}\left (\ket{H,0}\bra{H,0} + \ket{0,H}\bra{0,H} \right )  \nonumber \\
& & + \frac{r \epsilon_2^2 (1 - \epsilon_1^2)}{2}\left (\ket{V,0}\bra{V,0} + \ket{0,V}\bra{0,V} \right ) \nonumber \\
& & + (1 - r + r(1 - \epsilon_1^2)(1 - \epsilon_2^2))\ket{0,0}\bra{0,0},
\label{Eq: Appendix Beam splitter state}
\end{eqnarray}
where, e.g., $\ket{HV,0}$ denotes the case where both (distinguishable) photons leave the same beam splitter output port.

We now assume that the two photo detectors have the quantum efficiencies $\eta_1^2$ and $\eta_2^2$. The probability of detecting a coincidence event at $t=0$ will then be
\beq
P_D(0) = \frac{r \epsilon_1^2 \epsilon_2^2 \eta_1 \eta_2}{2}.
\eeq

The probability of getting a click in detector one at $t=0$ will be
\begin{eqnarray}
  P_{D1}(0) &=& r \eta_1^2 \left (\epsilon_1^2 \epsilon_2^2 + \frac{\epsilon_1^2 (1-\epsilon_2^2)}{2}  + \frac{\epsilon_2^2 (1-\epsilon_1^2)}{2}\right ) \nonumber \\
   &=& \frac{r \eta_1^2}{2} \left (2 \epsilon_1^2 \epsilon_2^2 + \epsilon_1^2 - \epsilon_1^2 \epsilon_2^2 + \epsilon_2^2 - \epsilon_1^2 \epsilon_2^2\right )\nonumber \\
   &=& \frac{r \eta_1^2}{2} \left ( \epsilon_1^2 + \epsilon_2^2 \right ).
\end{eqnarray}
The corresponding probability for a detection by detector 2 is obtained by the index permutation $1 \leftrightarrow 2$. Since the photons emanating from the source at a different times are uncorrelated, the probability to detect a photon at detector $j$ at the time $t=\tau$ are the same. Since the correlator we have used only measures if the two detection events are coincident (to within a preset time window), there are two ways of getting an event. Either detector 1 clicks at some point, and detector 2 clicks at time $t=\tau$, or vice versa. Thus, the probability of getting a coincidence for the time separation $\tau$ will be
\begin{eqnarray}
P_D(\tau) &=& 2 \cdot \frac{r^2 \eta_1^4 \eta_2^4}{4} \left ( \epsilon_1^2 + \epsilon_2^2 \right )^2 \nonumber \\
&=& \frac{r^2 \eta_1^4 \eta_2^4}{2} \left ( \epsilon_1^2 + \epsilon_2^2 \right )^2.
\end{eqnarray}
(If one detector had been designated the ``start'' detector, and the other the ``stop'' detector, and only ``start-stop'' events would have been recorded, the corresponding probability would have been halved.) We see that the expression is symmetric in the indices $j=1,2$ and we also see that it now depends on the pair production rate $r$ squared, so under our conditions, this probability is significantly smaller that the coincidence probability at time $t=0$.

\subsection{Theory, blinking}

To see the influence of blinking emitters, we shall investigate what happens if one arm is blocked, that is, only one photon in a pair will reach the beam splitter. This of course immediately rules out any coincidences at time $t=0$ so that $P_D'(0) = 0$.

Suppose we block the $H$-photon in arm 1. The state then becomes
\begin{eqnarray}
\hat{\rho}_b & = & r \ket{0,V}\bra{0,V} +(1 - r)\ket{0,0}\bra{0,0}.
\label{Eq: Blocked state}
\end{eqnarray}
After suffering attenuation it is transformed to
\begin{eqnarray}
\hat{\rho}_{ba} & = & r \epsilon_2^2 \ket{0,V}\bra{0,V}+(1 - r \epsilon_2^2)\ket{0,0}\bra{0,0}.
\label{Eq: Blocked attenuated state}
\end{eqnarray}
If the state is sent through the 50:50 beam splitter then the output becomes
\beq
\hat{\rho}_{bo}  =  \frac{r \epsilon_2^2}{2}\left (\ket{V,0}\bra{V,0} +  \ket{0,V}\bra{0,V} \right ) +(1 -  r \epsilon_2^2)\ket{0,0}\bra{0,0}.
\label{Eq: Blocked beam splitter state}
\eeq
The probability of getting a detection event in detector 1 at time $t=0$ becomes
\beq
\frac{r \epsilon_2^2 \eta_1^2}{2},
\eeq
and the corresponding probability for detector 2 is
\beq
\frac{r \epsilon_2^2 \eta_2^2}{2}.
\eeq
Thus, again because each photon pair generation event is independent, the probability for a coincidence at $t=\tau$ is
\beq
2 \cdot \frac{r^2 \epsilon_2^4 \eta_1^2 \eta_2^2}{4} = \frac{r^2 \epsilon_2^4 \eta_1^2 \eta_2^2}{2}.
\eeq
If instead arm 2 is blocked, the corresponding coincidence detection probability is obtained by the index $j$ substitution $1 \leftrightarrow 2$.

Assuming we now simulate the blinking of quantum emitters by randomly blocking each arm. If the frequency of blinking is much lower than the other time scales involved, such as the mean rate of photon pair production and the inverse of the preset time window defining ``coincidence'', then we will only have to consider four distinct situations: both arms are blocked (both emitters are in their off-state), one arm is blocked and the other is not (one emitter, either 1 or 2, are in their off-state), or none of the arms are blocked (both emitters are in their on-state). If the emitter duty cycle (or on-state probability) is denoted $\pi_j$, then to get a coincidence at $t=0$ it is necessary that both arms are open, and the corresponding probability/rate is
\beq
P_D'(0) = \frac{r \epsilon_1^2 \epsilon_2^2 \eta_1^2 \eta_2^2 \pi_1 \pi_2}{2}.
\eeq
We see that changing the attenuation $\epsilon_j^2$ of one arm has the same effect on this probability as changing the emitter duty cycle $\pi_j$.

To get a coincidence at $t=\tau$, however, it suffices that one arm is unblocked. We get three contributions to the coincidence probability
\begin{eqnarray}
P_D'(\tau) & = & \frac{r^2 \eta_1^2 \eta_2^2}{2} \left [ \pi_1 \pi_2 (\epsilon_1^2 + \epsilon_2^2)^2 \right. \nonumber \\
& & \left. + \pi_1(1-\pi_2) \epsilon_1^4 + \pi_2(1-\pi_1) \epsilon_2^4 \right] \nonumber \\
& = & \frac{r^2 \eta_1^2 \eta_2^2}{2} \left [ \pi_1 \epsilon_1^4 + \pi_2 \epsilon_2^4 + 2 \pi_1 \pi_2 \epsilon_1^2\epsilon_2^2  \right].
\label{Eq: Final coincidence at tau}
\end{eqnarray}
This equation clearly illustrates that now the duty cycle and the attenuation of the respective arms (emitters) do not enter on the same footing. The first two terms on the right hand side of the equation's last line decrease linearly with the duty cycle $\pi_j$, but quadratically with the attenuation $\epsilon_j^2$. Thus, if the duty cycle of the arms (emitters) is halved, the coincidence count rate is also halved. If, instead, the transmission in one arm is halved by adding linear loss, the coincidence count rate due to the first two terms in (\ref{Eq: Final coincidence at tau}) is reduced to one quarter. The reason for this is that the rightmost term on the right hand side of (\ref{Eq: Final coincidence at tau}) comes from coincidences where the two photons passed through separate arms. Thus both arms need to be open which explains the factor $\pi_1 \pi_2$, and the probability of having both photons transmitted is $\epsilon_1^2\epsilon_2^2 $. The two leftmost terms on the right hand side of (\ref{Eq: Final coincidence at tau}) comes from contributions where both photons passed through the same arm, but at different times. In this case it only matters that this arm is unblocked, and if it is, then the assumption of low frequency blinking assures that if the arm is unblocked for the first photon, then it remains unblocked also for the second photon. Thus the probability due to (un)blocking is $\pi_j$. However, when it comes to transmission due to linear losses, both photons need to pass through in order for a coincidence to be possible. The probability for this to happen is $\epsilon_j^4$.

\subsection{Comparison}

In the non-blinking case, the ratio between the coincidences at time $t=0$ and $t=\tau$ becomes
\beq
\frac{P_D(0)}{P_D(\tau)} = \frac{\epsilon_1^2 \epsilon_2^2 }{r\left (\epsilon_1^2 + \epsilon_2^2 \right )^2}.
\eeq
If we assume that $\epsilon_1 = \epsilon_2$, then the expression simplifies to $1/(4 r)$. If, to compare with quantum dots, we set $r=1$, then we get the simple ratio 1/4. The reason this result differ from the result 1/2 derived in our above calculations for quantum dots is that we have not assumed a start and a stop detector, so that the coincidence at different times ($t= \tau$) is effectively twice that one would get if one only counted the coincidence events in a certain order, e.g., detector 1 as the start signal and detector 2 as the stop. Thus, the ratio (apart from the obvious factor $r^{-1}$) is half that we would have gotten if we used a start-stop coincidence measurement technique.

The ratio between the coincidence rate at time $t=0$ and $t=\tau$ if we have blinking is
\beq
\frac{P_D'(0)}{P_D'(\tau)} = \frac{\epsilon_1^2 \epsilon_2^2 \pi_1 \pi_2}{r \left (\pi_1 \epsilon_1^4 + \pi_2 \epsilon_2^4 + 2 \pi_1 \pi_2 \epsilon_1^2 \epsilon_2^2 \right )}.
\eeq

Under the simplified assumption that $\epsilon_1 = \epsilon_2$ the expression reduces to
\beq
\frac{P_D'(0)}{P_D'(\tau)} = \frac{ \pi_1 \pi_2}{r \left (\pi_1 + \pi_2 + 2 \pi_1 \pi_2 \right )}.
\eeq
One sees that the numerator is proportional to the duty cycle squared, where as the denominator is proportional to the duty cycle. Thus, this ratio will go to zero as the duty cycle decreases. For, e.g., $\pi_1 = \pi_2 = 1/2$ one gets the ratio $1/(6 r)$, which is clearly smaller by a factor of 2/3 than the number $1/(4 r)$ one would have gotten for ``non-blinking'', perfectly distinguishable photon pairs. The factor 2/3 is the same reduction that we found in the analysis of blinking quantum dots. Hence, the two experimental situations are equivalent except for the small production rate $r$ for spontaneously generated photon pairs.

\bibliography{Blinking}

\end{document}